\documentclass[nohyperref]{article}

\usepackage{microtype}
\usepackage{graphicx}
\usepackage{subfigure}
\usepackage{enumitem}
\usepackage{booktabs} 
\usepackage{bm}
\usepackage{color}
\usepackage{hyperref}

\newcommand{\MusicCap}{MuLaMCap}
\newcommand{\EvalSet}{MusicCaps~\cite{musiclm2023}}
\newcommand{\MusicLM}{MusicLM~\cite{musiclm2023}}
\newcommand{\website}{https://google-research.github.io/noise2music}
\newcommand{\websiteDisplay}{google-research.github.io/noise2music}

\usepackage[preprint]{icml2023}

\usepackage{amsmath}
\usepackage{amssymb}
\usepackage{mathtools}
\usepackage{amsthm}

\usepackage[capitalize,noabbrev]{cleveref}

\theoremstyle{plain}

\theoremstyle{definition}

\theoremstyle{remark}

\usepackage[textsize=tiny]{todonotes}

\icmltitlerunning{Noise2Music}

\begin{document}

\twocolumn[
\icmltitle{Noise2Music: Text-conditioned Music Generation with Diffusion Models}



\icmlsetsymbol{equal}{*}
\icmlsetsymbol{core}{$\dagger$}

\begin{icmlauthorlist}
\icmlauthor{Qingqing Huang}{equal,g}
\icmlauthor{Daniel S. Park}{equal,g}
\icmlauthor{Tao Wang}{core,g}
\icmlauthor{Timo I. Denk}{core,g}
\icmlauthor{Andy Ly}{core,g}
\icmlauthor{Nanxin Chen}{core,g}
\icmlauthor{Zhengdong Zhang}{g}
\icmlauthor{Zhishuai Zhang}{g}
\icmlauthor{Jiahui Yu}{g}
\icmlauthor{Christian Frank}{g}
\icmlauthor{Jesse Engel}{g}
\icmlauthor{Quoc V. Le}{g}
\icmlauthor{William Chan}{w}
\icmlauthor{Zhifeng Chen}{w}
\icmlauthor{Wei Han}{core,g}
\end{icmlauthorlist}

\icmlaffiliation{g}{Google Research}
\icmlaffiliation{w}{Work done while at Google}

\icmlcorrespondingauthor{Qingqing Huang}{qqhuang@google.com}

\icmlkeywords{Generative models, Diffusion, Music generation}

\vskip 0.3in
]



\printAffiliationsAndNotice{\icmlEqualContribution\icmlCoreContributors} 

\begin{abstract}
We introduce Noise2Music, where a series of diffusion models is trained to generate high-quality 30-second music clips from text prompts. Two types of diffusion models, a generator model, which generates an intermediate representation conditioned on text, and a cascader model, which generates high-fidelity audio conditioned on the intermediate representation and possibly the text, are trained and utilized in succession to generate high-fidelity music. We explore two options for the intermediate representation, one using a spectrogram and the other using audio with lower fidelity. We find that the generated audio is not only able to faithfully reflect key elements of the text prompt such as genre, tempo, instruments, mood, and era, but goes beyond to ground fine-grained semantics of the prompt. Pretrained large language models play a key role in this story---they are used to generate paired text for the audio of the training set and to extract embeddings of the text prompts ingested by the diffusion models.

Generated examples: 

\href{\website}{\website}

\end{abstract}

\section{Introduction}

Deep neural networks have been shown to have remarkable generative ability. In this work, we explore the generative capability of deep models for audio.

We introduce Noise2Music, a diffusion-based \cite{sohl2015deep,song2019generative,ho2020denoising} method of generating music from text prompts and demonstrate its capability by generating 30-second long 24kHz music clips.

\textbf{Modeling:} We train a series of cascading diffusion models \cite{ho2022cascaded}, where the first model learns the generative task of producing a compressed representation of a 30-second waveform from a text prompt, and the second model learns to generate a 16kHz~waveform conditioned on the compressed representation and optionally the text prompt. We have investigated two options for the intermediate representation: a log-mel spectrogram, or a 3.2kHz~waveform. 1D~U-Nets are used for learning the noise vectors for the diffusion model. The diffusion models are conditioned on user prompts in the format of free-form text, which are encoded by a pre-trained language model (LM) and ingested by the 1D~U-Net layers via cross attention. A final super-resolution cascader is used to generate the 24kHz~audio from the 16kHz waveform.

\textbf{Data mining:} A large amount of training data is crucial for producing high-quality samples from a deep generative model. We employ a data mining pipeline to construct a large-scale training dataset of diverse music audio clips, each paired with multiple descriptive text labels. The text labels for the audio are generated by employing a pair of pre-trained deep models: first, we use a large language model to generate a large set of generic music descriptive sentences as caption candidates; we then use a pre-trained music-text joint embedding model to score each unlabeled music clip against all the caption candidates and select the captions with the highest similarity score as pseudo labels for the audio clip. We are able to annotate O(150K) hours of {audio sources} this way
to constitute our training data.

\textbf{\MusicCap:} As a by-product of this work, we introduce MuLan-LaMDA Music Caption dataset (\MusicCap), consisting of O(400K) music-text pairs obtained by annotating the music content from AudioSet by the process described above. Compared to the original AudioSet ontology, where 141 out of 632 label classes are music related, the captions in \MusicCap ~come from a large vocabulary consisting of 4 million music descriptive sentences and phrases, which have a much higher degree of diversity and granularity. We expect this dataset to be utilized for applications beyond sound classification, e.g., music captioning, retrieval or generation.

\textbf{Evaluation:} We measure the quality of our text conditioned music generation model with two metrics: the Fr\'echet Audio Distance (FAD)~\cite{kilgour2018fr} which measures how the quality of generated audio clips compare to that of two benchmark datasets, e.g. the music split of AudioSet~\cite{gemmeke2017audio} and MagnaTagATune~\cite{law2009evaluation}; and the MuLan similarity score~\cite{mulan2022} which measures the semantic alignment between text prompts and the corresponding generated audio clips. 

\textbf{Generative ability:} Our models demonstrate that they can go beyond simple music attribute conditioning, e.g., genre, instrument, era, and are able to handle complex and fine-grained semantics which can reflect soft attributes such as atmosphere, feeling or activity. This is achieved by constructing a training dataset that not only relies on the metadata tags, but that also leverages the pre-trained music-text joint embedding model to ground the semantics to audio features. Cherry-picked examples of music generated from text prompts can be found in  \href{\website#table-1}{\websiteDisplay\#table-1} and \href{\website#table-4}{\websiteDisplay\#table-4}.

\section{Related Work}
\noindent\textbf{Generative models:}
Deep generative models have a long and successful history in a wide range of domains. More recently, a significant amount of effort has been focused toward scaling up the dataset size for training models that can produce extremely high quality samples. Here we compile an incomplete list of such recent developments in text~\cite{brown2020language,thoppilan2022lamda}, speech~\cite{wang2018style, chen2021wavegrad,audiolm}, images~\cite{ramesh2022hierarchical,saharia2022photorealistic,yu2022scaling}, and audio~\cite{briot2021artificial,dhariwal2020jukebox,mubert,kreuk2022audiogen}.

\noindent\textbf{Diffusion models:}
Diffusion models, introduced in \cite{sohl2015deep,song2019generative,ho2020denoising} have shown the capability to generate high quality images \cite{ho2020denoising,ho2022cascaded}, audio \cite{yang2022diffsound, popov2021grad} and video \cite{ho2022video,ho2022imagen}. Cascaded diffusion models \cite{ho2022cascaded,saharia2022photorealistic}, which uses a series of diffusion models to generate a low-fidelity image and refine the image in succession to produce a high-fidelity image, has been adapted to audio in this work.

\noindent\textbf{Audio generation:}
Various methods have been employed to generate audio conditioned on external input. Some relevant examples are provided in the context of the text-to-audio task, in which text-conditioned spectrogram generation and spectrogram-conditioned audio has been intensively studied \cite{popov2021grad, chen2021wavegrad, kong2021diffwave, wu2021otts, chen2022infergrad}. Restricting our attention to audio generation based on descriptive text, text conditioned general sound event generation has been approached with auto-regressive methods by AudioGen~\cite{kreuk2022audiogen} as well as diffusion-based methods that operate on discrete audio codes by DiffSound~\cite{yang2022diffsound}. If we narrow our scope to music generation, Jukebox~\cite{dhariwal2020jukebox}, Mubert~\cite{mubert}, and \MusicLM~have taken an auto-regressive approach, while Riffusion~\cite{Forsgren_Martiros_2022} employed diffusion for spectrogram generation.

\noindent\textbf{Conditional signals in audio generation:}
Broadly speaking, two approaches have been taken on how the conditional signal, which steers the model to generate a specific style of music, is parameterized and communicated to an audio generation model. One approach is to project the signal to a pre-defined, interpretable embedding space---Jukebox~\cite{dhariwal2020jukebox} relies on a fixed vocabulary of artists and genres mined from the training data to condition the decoder, while Mubert~\cite{mubert} matches the user prompt to a set of tags in a predefined vocabulary. The other, taken by works such as AudioGen~\cite{kreuk2022audiogen} and \MusicLM~is to use a pre-trained text encoder to encode arbitrary user prompts.

\noindent\textbf{Authors' Note:}
During the completion of this work, concurrent research which has overlap with this work has appeared \cite{schneider2023}.

\section{Methods}

\subsection{Diffusion models}
\label{ss:diffusion-models}

Diffusion models \cite{sohl2015deep,song2019generative,ho2020denoising} are powerful generative models that generate a sample by iteratively denoising random noise. Here we review the minimal amount of information on diffusion models required for understanding our work. More details can be found in the supplementary material.

The input to a diffusion model, which we consider to be a generative model of some sample space, is the conditioning signal $\mathbf{c}$, a randomly sampled time step $t$ and a sample $\mathbf{x}_t$ obtained by corrupting the original sample $\mathbf{x}$ via a Gaussian diffusion process with a noise schedule parameterized by the standard deviation $\sigma_t$ of the noise at time $t$. The range of time $t$ is set to be $[0, 1]$, from which it is uniformly sampled during training, and the diffusion is viewed to progress in the direction of increasing time. The dynamics of Gaussian diffusion are well understood---the distribution of $\mathbf{x}_t$ is completely parameterized by a single noise vector $\bm{\epsilon}$ that belongs to a standard normal distribution, as $\mathbf{x}_t$ maybe written as a function of the original sample, the deterministic noise schedule, and the noise vector $\bm{\epsilon}$, i.e., $\mathbf{x}_t(\mathbf{x}, \sigma, \bm{\epsilon})$, where it should be understood that $\sigma$ is used to denote the entire noise schedule. The model $\bm{\epsilon}_\theta$ is trained to identify the noise vector given this input. The diffusion loss can be written as
\begin{equation}
\mathbb{E}_{\mathbf{x}, \mathbf{c}, \bm{\epsilon}, t}
\left[ w_t \Vert  \bm{\epsilon}_\theta (\mathbf{x}_t, \mathbf{c}, t)- \bm{\epsilon} \Vert^2 \right] \,,
\end{equation}
where $w_t$ is a fixed weight function of choice.

Inference is carried out by taking random noise at time $t=1$ and denoising it by utilizing the noise predictions given by the model. We use ancestral (or DDPM) sampling \cite{ho2020denoising}, which provides a flexible framework for inference allowing multiple parameters that can affect the quality of the generated sample. First, the level of stochasticity of the denoising process can be controlled by varying the stochasticity parameter $\gamma$ of the sampler. Also, an arbitrary denoising schedule can be used, where one may choose an arbitrary partition of the interval $0 = t_0 < \cdots < t_n =1$ to discretize the denoising steps.

Thus a variety of choices present themselves when one wishes to train a diffusion model. We utilize multiple options with respect to the following elements, further details of which can be found in the supplementary material:
\begin{itemize}[itemsep=1pt]
\item Loss weight ($w_t$): simplified weight $w_t =1$ \cite{ho2020denoising} and sigma weight $w_t = \sigma_t^2$
\item Variance schedule: linear \cite{ho2020denoising} and cosine \cite{nichol2021improved} schedules
\item Stochasticity parameter: $\gamma= 0$ or $1$
\item Denoising step schedule
\end{itemize}

\textbf{Classifier-free guidance (CFG):} CFG \cite{ho2022classifier} is a method for improving the alignment between generated samples and conditional inputs. The conditional input of a portion of the training samples in each training batch are hidden from the network during training, enabling the network to learn how to predict the noise vector unconditionally and conditionally. At inference, the noise vector with and without the conditional input are computed, and the final noise vector applied is set to $w \bm{\epsilon}_\theta(\mathbf{x}_t, \mathbf{c}) + (1 - w) \bm{\epsilon}_\theta(\mathbf{x}_t, \cdot) $ with $w > 1$. Dynamic clipping \cite{saharia2022photorealistic} is applied to avoid over-saturation due to CFG.

\subsection{Architecture}

\begin{figure*}[ht]
\vskip 0.05in
\begin{center}
\centerline{\includegraphics[width=1.7\columnwidth]{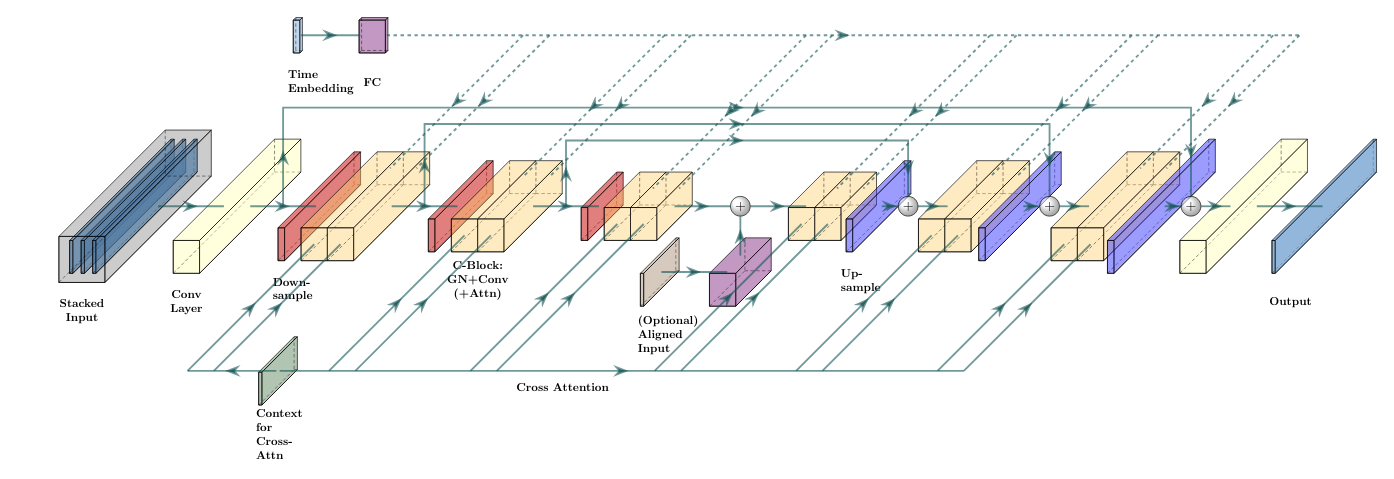}}
\vskip -0.05in
\caption{U-Net architecture used for the diffusion models. A series of down-sampling, then up-sampling blocks is applied to the main input sequence of length $T$ to produce an output sequence of length $T$. The outputs of the down-sampling blocks are added to the outputs of up-sampling blocks via residual connections. There are four modes of inputs to the model. The first is the (possibly stacked) main input of sequence length $T$, entering on the left-hand side of the diagram. $T$ is the target sequence length. Second, there is a time embedding vector. Third, there can be a text embedding sequence that can be attended to by the down/up-sampling blocks via cross attention. Lastly, there can be input of length $T/C$ that is aligned with the sequence of length $T$ with compression rate $C$.}
\label{U-Net}
\end{center}
\vskip -0.25in
\end{figure*}

We deploy the 1D Efficient U-Net, a one-dimension version of the Efficient U-Net introduced in \cite{saharia2022photorealistic}, for the diffusion model. The U-Net model, depicted in \cref{U-Net}, consists of a series of down-sampling and up-sampling blocks which are connected by residual connections. A down/up-sampling block consists of a down/up-sampling layer followed by a series of blocks obtained by composing 1D convolutional layers, self/cross-attention layers and combine layers. The combine layer enables a single vector to interact with a sequence of vectors, where the single vector is used to produce a channel-wise scaling and bias. These blocks closely follow the structure of the blocks of the efficient U-Nets constructed in \cite{saharia2022photorealistic}, with the two-dimensional convolutions replaced by their one-dimensional counterparts. The exact structure of the blocks are further elaborated in the supplementary material.

There are four possible routes of entry to the model. The stacked input and output both consist of sequences of some length $T$, while the diffusion time $t$ is encoded into a single time embedding vector and interacts with the model through the aforementioned combine layers within the down and up-sampling blocks. Given that we would like to produce a sequence of length $T$, the noisy sample $\mathbf{x}_t$ is always part of the stacked input on the left-most side of the figure, while the output is interpreted as the noise prediction $\bm\epsilon$. For the cascading models, the low-fidelity audio on which the model is conditioned on can be up-sampled and stacked. Meanwhile, a sequence of vectors with an arbitrary length may interact with the blocks through cross-attention. This is the route through which the text prompts are fed into the model. There is also room for the model to be conditioned on an aligned, but compressed representation of the sequence by addition at the bottom of the ``U" of the U-Net.

\subsection{Cascaded diffusion}

We train two kinds of diffusion models in this work to produce high-quality 30-second audio from text prompts. Following \cite{ho2022cascaded}, we train generator models that generate some intermediate representation of the final audio conditioned on a text prompt, and cascader models that produce the final audio based on the intermediate representation. For the intermediate representation, we consider both low-fidelity audio and spectrograms.

\subsubsection{Waveform Model}

\textbf{Generator Model:} The generator model generates 3.2kHz audio that is conditioned on the text input. A sequence of vectors derived from the text input is produced and fed into the network as a cross-attention sequence.

\textbf{Cascader Model:} The cascader model generates 16kHz audio that is conditioned on both the text prompt and the low-fidelity audio generated by the generator model based on the text prompt. The text conditioning takes place via cross attention. Meanwhile, the low-fidelity audio is up-sampled and stacked with $\mathbf{x}_t$ and fed into the model. The up-sampling is done by applying fast Fourier transform (FFT) to the low-fi audio sequence and then applying inverse FFT to obtain the high-fi audio from the low-fi Fourier coefficients.

\subsubsection{Spectrogram Model}

\textbf{Generator Model:} This model generates a log-mel spectrogram conditioned on the text input. The spectrgram has 80 channels and a frequency of 100 features per second. The input and output sequences now have a channel dimension in addition to the sequence dimension. The pixel values of the log-mel spectrogram are normalized to lie within $[-1, 1]$. Text conditioning is achieved through cross attention.

\textbf{Vocoder Model:} The vocoder model generates 16kHz audio that is conditioned only on the spectrogram, which is treated as aligned input. The down and up-sampling rates of the U-Net model are tuned to achieve the compression rate of the spectrogram against the audio.

\subsubsection{Super-resolution cascader}

A final light-weight cascader is used to generate 24kHz audio from the 16kHz waveform produced by either model. The 16kHz audio is up-sampled and stacked with $\mathbf{x}_t$ as input to the model. Text conditioning is not used for this model.

\subsection{Text understanding}

It has been shown in the context of text-to-image diffusion models \cite{saharia2022photorealistic,rombach2021highresolution} that powerful text encoders are able to capture the complexity and compositionality of music descriptive text prompts. We adopt the T5 encoder \cite{raffel2020exploring} and use the non-pooled token embedding sequence to condition the diffusion models. A thorough comparison with alternative contextual signals such as embeddings from different large language models, or a single vector embedding derived from CLIP-like \cite{radford2021learning} text encoders trained on music-text pairs \cite{mulan2022,ilaria-contrastive-audio} is beyond the scope of this work.

\subsection{Pseudo labeling for music audio}
Having large scale training data is a necessary component for ensuring the quality of generative deep neural networks. For example, Imagen \cite{saharia2022photorealistic} was trained on O(1B) image-text pairs.
Despite the fact that music content is widely available, high quality paired music-text data is scarce, especially in the case of free-form text that describes the music attributes beyond high-level metadata such as title, artist name, album name, and release year.

To generate such music-text pairs, we take a pseudo-labeling approach via leveraging MuLan~\cite{mulan2022}, a pre-trained text and music audio joint embedding model, together with LaMDA~\cite{thoppilan2022lamda}, a pre-trained large language model, to assign pseudo labels with fine-grained semantic to unlabeled music audio clips.

We first curate several music caption vocabulary sets, each consisting of a large list of music descriptive texts. As demonstrated below, these texts vastly differ from the captions from the label classes in standard music classification benchmarks, e.g., MagnaTagATune, FMA, and AudioSet, in their scale and the fine-grained semantic granularity. We consider the following three caption vocabularies:

\textbf{LaMDA-LF}: We prime the large language model LaMDA to describe a list of 150k~popular songs provided the song title and artist names. The precise prompt template is provided in the supplementary material. We then process the LaMDA responses into 4~million clean long-form sentences that are likely to be describing music.
We use LaMDA as our LM of choice because it is trained for dialogue applications, and expect the generated text to be closer to user prompts for generating music. 

\textbf{Rater-LF}: We obtain 10,028 rater written captions from \EvalSet, and split each caption into individual sentences. This produces 35,333 music-describing long-form sentences.

\textbf{Rater-SF}: From the same evaluation set above, we collect all the short-form music aspect tags written by the raters, which amounts to a vocabulary of size 23,906.

Examples of the caption vocabulary are presented in \cref{table:caption-vocab}.

\begin{table}
\caption{Caption vocabulary examples.}
\vskip 0.05in
  \label{table:caption-vocab}
  \centering
  \resizebox{0.95\columnwidth}{!}{
  \begin{tabular}{ll}
    \toprule
    {\bf Vocabulary} & {\bf Examples}  \\
    \midrule
    \begin{tabular}{@{}l@{}}
    LaMDA-LF \\
    (4M)
    \end{tabular}
    &
    \begin{tabular}{@{}l@{}}
    ``A light, atmospheric drum groove provides a tropical feel.", \\
    ``A light EDM drumbeat carries a bass guitar, strings, \\
    ~a simple piano, and percussion in the background." \\
    \end{tabular} \\
    \midrule
    \begin{tabular}{@{}l@{}}
    Rater-LF \\
    (35k)
    \end{tabular} &
    \begin{tabular}{@{}l@{}}
    ``A Scottish tenor drum plays a marching beat."\\
    ``A bass guitar with a punchy sound contrasts the guitar." \\
    \end{tabular} \\
    \midrule
    \begin{tabular}{@{}l@{}}
    Rater-SF \\
    (24k)
    \end{tabular} &
    \begin{tabular}{@{}l@{}}
    ``50's pop", ``wide passionate male vocal", ``vintage vibes",\\
     ``patriotic mood", ``vivacious cello", ``exercise music" \\
    \end{tabular} \\
    \bottomrule
  \end{tabular}
  }
\vskip-0.1in
\end{table}

We use the MuLan model as a zero-shot music classifier to assign captions from the vocabulary to unlabeled audio clips.
MuLan consists of a text encoder and an audio encoder, which are trained on a large amount of highly noisy text-music pairs with a contrastive learning scheme. Similar to how CLIP \cite{radford2021learning} co-embeds image and text, a 10-second long music audio clip and a sentence that describes the music are placed closely in the same semantic embedding space learned by MuLan.
For each audio clip, we compute its audio embedding by first segmenting the clip into non-overlapping 10-second windows, and computing the average of the MuLan audio embeddings of each window. The text embeddings of all the candidate captions in the vocabulary are also computed. The top $K$ captions that are closest to the audio in the embedding space are selected for each clip.  
We compute the frequency counts of the captions among all clips, and for each clip further sample $K'$ out of the $K$ captions with probabilities inverse to the frequency counts.
This last sampling step serves to balance the label distribution and increase the diversity of the captions. We use $K=10$ and $K'=3$.

As a warm up for pseudo-labeling our large training set, we produce \MusicCap, a music captioning dataset derived from the AudioSet \cite{gemmeke2017audio}, by applying this pseudo-labeling method to 388,262 / 4,497 examples from the AudioSet train / test sets which have labels in the music subtree of AudioSet ontology. Each 10-second audio with music content is associated with 3~captions from the LaMDA-LF vocabulary, 3~captions from Rater-LF, and 6~short form captions from Rater-SF.

\subsection{Training data mining}
To assemble a large-scale collection of audio-text pairs, we collect approximately 6.8M {music audio source files.}
From each {soundtrack}, we extract six non-overlapping 30-second clips. This amounts to nearly 340k~hours of music. Audio is sampled at 24kHz for training the super-resoluton model and 16kHz for training all other models.

For each {soundtrack}, we consider three types of noisy text labels---the song title, named entity tags associated with the {soundtrack} (e.g., genre, artist name, instrument), and the pseudo labels. We use three pseudo labels from the LaMDA-LF vocabulary, and six pseudo labels from the Rater-SF vocabulary.
The pseudo labels from LaMDA-LF and Rater-SF provide complementary information to the named entity tags. Compared to the objective and high-level tags, the pseudo labels include subjective descriptions related to activity (``music for highway driving") and mood (``a laid back feel"), and also include compositional elements with fine-grained semantics. 
Since we evaluate our model on \EvalSet~from which the sentences of Rater-LF were derived, we exclude any pseudo labels from the Rater-LF vocabulary from our training data.

We include a small amount of high-quality audio to the large pseudo-labeled training set. The audio is taken from a subset of music tracks, which does not require attribution, from an {internally}  {maintained} music library. The music tracks are segmented to non-overlapping 30-second clips, while the metadata of the tracks are concatenated to form the text prompt of the audio. This contributes O(300) hours of annotated audio to our training data.

\section{Experiments and Results}

\subsection{Model training details}
We train four 1D~U-Net models, the waveform generator and cascader, and the spectrogram generator and vocoder for this work. We have summarized some basic information about the models in \cref{t:model-specs}, while we relegate further details about the models to the supplementary material. We note that we found the sigma-weighted loss, which weighs the loss more heavily on the ``back end" of the denoising schedule, crucial for convergence of the spectrogram generator.

\begin{table}
\caption{Models trained in this work. The token length refers to the token length of the text prompts at training time.}
\vskip 0.05in
  \label{t:model-specs}
  \centering
  \resizebox{0.95\columnwidth}{!}{
  \begin{tabular}{lccccc}
    \toprule
    {\bf Model}
    & {\bf \# Params}
    & {\bf \# Training}
    & {\bf Token}
    & {\bf Loss}
    & {\bf Noise} \\[-1pt]
    & 
    & {\bf steps}
    & {\bf length}
    & {\bf weight} 
    & {\bf schedule} \\
    \midrule
    Waveform generator & 724M & 1.6M & 64 & Simplified & Cosine \\
    Waveform cascader & 487M & 460k & 64 & Simplified & Linear \\
    \midrule
    Spectrogram generator & 745M  & 1.8M & 96 & Sigma & Linear \\
    Spectrogram vocoder & 25.7M & 840k & - & Simplified & Linear \\
    \midrule
    Super-resolution cascader & 81M & 270k & - & Simplified & Linear \\
    \bottomrule
  \end{tabular}
  }
\vskip-0.2in
\end{table}

All the models, with the exception of the vocoder, are trained on audio-text pairs, while the vocoder is only trained on audio. For each audio sample, a text batch is formed. The three long prompts constitute three independent elements of the text batch, while the shorter prompts are concatenated, then segmented into a set token length reported in \cref{t:model-specs} and added to the text batch. For each audio clip, a random element of the corresponding text batch is selected at training time and fed to the model as the paired text to the audio.

The models are trained with Adam optimization with $\beta_1=0.9$ and $\beta_2=0.999$. A cosine learning rate schedule with the end point set to 2.5 M steps is used with peak learning rate 1e-4 and 10k warm-up steps. An exponential moving average (EMA) of the model parameters are taken with decay rate 0.9999 and used at inference time. The super-resolution cascader is trained with batch size 4096, while all other models use batch size 2048. To apply CFG at inference time, we occlude the text prompts for 10\% of the samples in each training batch. For these samples, the output of the cross attention layers are set to zero.

While the generator models use self-attention, the cascaders and vocoder do not. Thus while we need to train the generator models on the entire 30-second representation of the audio, the cascader and vocoder models are trained on 3 to 4-second randomly sampled snippets.

Following \cite{ho2022cascaded}, two augmentations are applied at training time for the cascader/vocoder models. One is to randomly corrupt the conditioning low-fidelity audio or the spectrogram input by applying diffusion noise. To do so, a random diffusion time is chosen within $[0, t_\text{max}]$ and applied to the intermediate representation of the audio, i.e., the up-sampled low-fi audio or the spectrogram. For the cascader $t_\text{max}$ is set to 0.5 while for the vocoder and super-resolution cascader it is set to 1.0. The other is blur augmentation. For the cascader model, a 1D blur kernel of size $10$ is used with a Gaussian blur kernel whose standard deviation ranges from $0.1$ to $5.0$. For the vocoder model, a 2D 5x5 blur kernel is applied with the standard deviation ranging from 0.2 to 1.0.

\subsection{Model inference and serving}

\subsubsection{Model inference}

\begin{table}
\caption{Inference parameters for the models used in this work.}
\vskip 0.05in
  \label{t:model-inference-params}
  \centering
  \resizebox{0.9\columnwidth}{!}{
  \begin{tabular}{lccc}
    \toprule
    {\bf Model}
    & {\bf Denoising}
    & {\bf Stochasticity}
    & {\bf CFG scale}\\[-1pt]
    & {\bf step schedule}
    & {\bf parameter}
    &\\
    \midrule
    Waveform generator & Front-heavy & 0 & 10.0\\
    Waveform cascader & Front-heavy & 1 & 5.0 \\
    \midrule
    Spectrogram generator & Back-heavy & 0 & 5.0 \\
    Spectrogram vocoder & Front-heavy & 0 & N/A \\
    \midrule
    Super-resolution cascader & Front-heavy & 0 & N/A \\
    \bottomrule
  \end{tabular}
  }
\vskip-0.2in
\end{table}

We adjust three inference hyperparameters, the denoising schedule, the stochasticity parameter, and the CFG scale. The parameters used for each model are listed in \cref{t:model-inference-params}.

We parameterize the denoising step schedule by the time step sizes $[\delta_1, \cdots, \delta_N]$ that translate into denoising steps introduced in section \ref{ss:diffusion-models} via accumulation: $t_n = \sum_{i=1}^n \delta_n$. The inference cost is proportional to the number of time-steps. Thus optimizing the time step schedule with a fixed inference cost amounts to distributing a fixed number of time steps that add up to the total time, 1. The parameter space for the denoising step schedule is extremely large. Nevertheless, we experiment with three different kinds of schedules we denote ``front-heavy," ``uniform," and ``back-heavy." The front-heavy schedule allots many steps to the ``front" of the schedule near $t=0$ whereas the ``back-heavy" schedule expends more steps near $t=1$. The uniform schedule uses evenly-spaced time steps. The exact schedules used are produced in the supplementary material.

\subsubsection{Model serving}
We serve the models on Google Cloud TPU~V4, where each service request generates four 30-second music clips. We apply GSPMD \cite{gspmd} to partition the model on four TPU~V4 devices, reducing the serving time by more than 50\%. \cref{table:decode-latency} shows the inference time cost when the model is served on four TPU V4 to produce four samples.

\begin{table}
\caption{Inference time cost on four TPU~V4 for four samples.}
\vskip 0.05in
  \label{table:decode-latency}
  \centering
  \resizebox{0.8\columnwidth}{!}{
  \begin{tabular}{lrrr}
    \toprule
    {\bf Model} & {\bf time/step (ms)} & {\bf steps} & {\bf time (s)}  \\
    \midrule
    Waveform generator & 25.0 & 1000 & 25.0 \\
    Waveform cascader & 75.0 & 800 & 60.0 \\
    \midrule
    Spectrogram generator & 8.3 & 1000 & 8.3 \\
    Spectrogram vocoder & 29.9 & 100 & 0.3 \\
    \midrule
    Super-resolution cascader & 71.7 & 800 & 57.3 \\
    \bottomrule
  \end{tabular}
  }
\vskip-0.2in
\end{table}

\subsection{Evaluation}

\subsubsection{Parameter selection for the models}
\label{ss:model-parameter-selection}

Model parameters, including the architecture, training hyperparameters, checkpoints and inference parameters are selected in a heuristic fashion. A small set of dev prompts, independent of the prompts in any of the evaluation sets presented, are devised by the authors, which are used to generate audio from the trained models. Model parameters are selected based on the quality of the generation results, evaluated according the judgement of the authors, as well as practical limitations such as the availability of computational resources and time. Evaluations are conducted on 16kHz waveforms---the super-resolution cascader is not utilized to generate waveforms for producing evalution metrics.

\subsubsection{Evaluation metrics} 
We measure the quality of our text conditioned music generation model with two kinds of metrics: the Fr\'echet Audio Distance (FAD)~\cite{kilgour2018fr} and the MuLan similarity score~\cite{mulan2022}. 

FAD measures how the audio quality of the generated audio examples compare to that of a set of reference audio clips.
In particular, an audio encoder is used to compute the audio embeddings of both the set of generated audio examples and the background audio clips in the evaluation dataset. Assuming that the distribution of the embeddings from each set are Gaussian, and the Fre\'echet distance between the two distributions are computed from the mean embedding vectors and the correlation matrix of the two sets.

Three audio encoders are utilized for computing the FAD metric:
a VGG\footnote{\href{https://tfhub.dev/google/vggish/1}{tfhub.dev/google/vggish/1}} audio event embedding model \cite{cnn-for-audio-cls} trained on YouTube-8M \cite{yt8m-data};
the Trill \cite{trill} model\footnote{\href{https://tfhub.dev/google/nonsemantic-speech-benchmark/trill/3}{tfhub.dev/google/nonsemantic-speech-benchmark/trill/3}}, a convolutional speech representation learning model trained on speech containing clips from AudioSet; and the MuLan audio encoder. VGG and Trill produce frame-wise embeddings while MuLan's embeddings are clip-wise. Since the audio encoders are trained on different datasets and tasks, FAD computed with those audio representations focus on different aspects of the audio. We hypothesize that FAD\textsubscript{VGG} evaluates the general audio quality, FAD\textsubscript{Trill} is more indicative of the vocal quality, and FAD\textsubscript{Mulan} captures global musical semantics. 

The contrastive model MuLan provides us a way to quantify the similarity between audio-text pairs as well as audio-audio pairs. %
For a given text-audio or audio-audio pair, we define the MuLan similarity as the cosine similarity between the MuLan embeddings of the two entities. For a given evaluation set of music-text pairs, we compute the average similarity between the audio generated from the text prompts of the dataset and either the text or the ground truth audio associated to the text.
As a reference, we also compute the average MuLan similarity of the evaluation set against the ground truth audio, as well as a ``random" audio pairing obtained by shuffling the ground truth audio.

\subsubsection{Evaluation datasets}

We report the FAD and MuLan similarity with respect to the following three datasets consisting of text-music pairs.

First, we re-purpose the audio tagging benchmark MagnaTagATune (MTAT)~\cite{law2009evaluation} to evaluate the 29-second long music clips generated by our models.
MTAT contains 25,863 music clips, 21,638 of which are associated with multiple tags from a vocabulary of 188 music tags. We only utilize these 21,638 examples for evaluation, for each of which we concatenate the music tags with into a single string that we use as the associated text prompt. During evaluation, we generate a single 29-second long audio clip for the prompt associated with each of the 21,638 examples.

Second, we use AudioSet-Music-Eval, the music portion of AudioSet~\cite{gemmeke2017audio}. There are 1,482 music related examples in the evaluation split of AudioSet, where each 10-second clip is associated with labels from the non-trivial part of the music subtree of AudioSet ontology. For each example in this set, we use the concatenated labels as the text prompt to generate a 30-second long clip, the middle 10-second portion of which is used for evaluation. 

Lastly, we evaluate on \EvalSet~which consists of 5.5K 10-second clips from AudioSet paired with rater written captions. We use the rater captions as text prompts, and report the metrics with the middle 10-second excerpts of the generated audio samples.

\subsection{Evaluation results}
In \cref{table:eval-FAD}, we report the FAD of our models on the three evaluation datasets, and compare them with baseline models from Riffusion\footnote{We query Mubert API at \href{https://github.com/MubertAI}{github.com/MubertAI} as of Dec~24, 2022 to generate 10-second audio clips given the text prompts in the evaluation datasets.} and Mubert\footnote{We ran inference with riffusion-model-v1 provided by \href{https://github.com/riffusion/riffusion-app}{github.com/riffusion/riffusion-app} as of Dec~24, 2022 to generate 10-second audio clips.}. 
In \cref{table:eval-Mulan}, we report the average audio-text and audio-audio MuLan similarity scores between the generated audio and the evaluation datasets. We also include the metrics computed for the ground truth audio, as well as the shuffled ground truth.

The evaluation metrics should be interpreted with care, since our result has potential advantages over the baselines presented. First, there is a possibility that our training data distribution is closer to the evaluation datasets compared to the baselines. Also, one may suspect that the MuLan-based metrics might be biased towards our models, since the MuLan model has been used to pseudo-label our data. The reader should thus be cautious to draw conclusions about the effectiveness of the methods used in this paper compared to those of the baselines based on these metrics. The metrics, however, are indeed representative of the performance of the trained models themselves in the AudioSet domain and provides a quantitative measure of final model performance. This ideally should hold true for the MuLan-based metrics as well, if we assume that the MuLan model, much like CLIP \cite{radford2021learning}, has learned an un-biased, faithful representation of text and audio data.

\begin{table}[t!]
\vskip -0.05in
\caption{The FAD between the reference dataset audio and the generated audio with prompts from the reference dataset. Three audio encoders, VGG, Trill and MuLan have been used to measure FAD. A lower value indicates better proximity of quality.}
  \label{table:eval-FAD}
  \vskip 0.05in
  \centering
  \resizebox{0.95\columnwidth}{!}{\begin{tabular}{lrrr}
    \toprule
    {\bf Dataset/Model} & {\bf FAD\textsubscript{VGG}}  & {\bf FAD\textsubscript{Trill}} &  {\bf FAD\textsubscript{MuLan}}  
    \\
    \midrule
    \multicolumn{4}{l}{\bf \EvalSet} \\[4pt]
    ~~~~Riffusion~\cite{Forsgren_Martiros_2022} &
    13.371 & 0.763 & 0.487 \\
    ~~~~Mubert~\cite{mubert} &
    9.620 & 0.449 & 0.366 \\
    ~~~~\MusicLM &
    4.0~~~~ & 0.44~~ & -  \\[4pt]
    ~~~~{Noise2Music Waveform} &
    \bf 2.134 & \bf 0.405 & \bf 0.110 \\
    ~~~~{Noise2Music Spectrogram}    & 3.840  & 0.474  & 0.180 \\
    \midrule
    \multicolumn{4}{l}{\bf AudioSet-Music-Eval} \\[4pt]
    ~~~~{Noise2Music Waveform}
    & 2.240  & 0.252  & 0.193 \\
    ~~~~{Noise2Music Spectrogram}    & 3.498  & 0.323  & 0.276  \\
    \midrule
    \multicolumn{4}{l}{\bf MagnaTagATune} \\[4pt]
    ~~~~{Noise2Music Waveform}
    & 3.554  & 0.352 & 0.235 \\
    ~~~~{Noise2Music Spectrogram}    & 5.553  & 0.419  & 0.346  \\
    \bottomrule
  \end{tabular}}
\vskip-0.15in
\end{table}

\begin{table}[t!]
\vskip -0.05in
\caption{The average MuLan similarity between the generated audio and either
the text prompt or the ground truth audio for each evaluation set. A higher value
indicates better semantic alignment.}
  \label{table:eval-Mulan}
  \vskip 0.05in
  \centering
  \resizebox{0.95\columnwidth}{!}{\begin{tabular}{lrr}
    \toprule
    {\bf Dataset/Model} &
    \bf audio $\leftrightarrow$ gt-text  &
    \bf audio $\leftrightarrow$ gt-audio  \\
    \midrule
    \multicolumn{3}{l}{\bf \EvalSet} \\[4pt]
    ~~~~Ground Truth Audio & 0.452 & (1.000) \\
    ~~~~Randomly Shuffled Audio & 0.248 &  0.278 \\[4pt]
    ~~~~Riffusion~\cite{Forsgren_Martiros_2022} &
    0.342 & 0.312 \\
    ~~~~Mubert~\cite{mubert} &
    0.323 & 0.280 \\
    ~~~~\MusicLM & \bf 0.51~~ & -  \\[4pt]
    ~~~~{Noise2Music Waveform}
    & 0.478 & \bf 0.489  \\
    ~~~~{Noise2Music Spectrogram}    & 0.434  & 0.464  \\
    \midrule
    \multicolumn{3}{l}{\bf AudioSet-Music-Eval} \\[4pt]
    ~~~~Ground Truth Audio & 0.470 & (1.000) \\
    ~~~~Randomly Shuffled Audio & 0.274 & 0.265 \\[4pt]
    ~~~~{Noise2Music Waveform}
    & \bf 0.563  & \bf 0.429  \\
    ~~~~{Noise2Music Spectrogram}    & 0.490  & 0.389  \\
    \midrule
    \multicolumn{3}{l}{\bf MagnaTagATune} \\[4pt]
    ~~~~Ground Truth Audio & 0.498 & (1.000) \\
    ~~~~Randomly Shuffled Audio & 0.277 & 0.315 \\[4pt]
    ~~~~{Noise2Music Waveform}
    & \bf 0.518 & \bf 0.479 \\
    ~~~~{Noise2Music Spectrogram}    & 0.459  & 0.444  \\
    \bottomrule
  \end{tabular}}
\vskip-0.1in
\end{table}

We also conduct human listening tests to measure the semantic alignment, whose setup is identical to that used in \cite{musiclm2023} with now five sources, listed in Table~\ref{table:human-study}. The participants of the test are presented with a text caption from the MusicCaps evaluation set and 10-second clips from two different sources, then asked which clip is better described by the text of the caption on a 5-point Likert scale. We collect 3k ratings, with each source involved in 1.2k pair-wise comparisons. The the head-to-head comparisons between each pair of sources is evenly distributed.
In Table~\ref{table:human-study}, we report the total number of ``wins'' each model achieved among the 1.2k comparisons it has been subject to. Our waveform model shows comparable performance to MusicLM, while being behind the ground truth audio.
\begin{table}[t!]
\caption{The number of wins in pair-wise comparisons of the human listening study. Higher indicates better semantic alignment.}
  \label{table:human-study}
  \vskip 0.05in
  \centering
  \resizebox{0.95\columnwidth}{!}{\begin{tabular}{ccccc}
    \toprule
    \textbf{MusicCaps} & \textbf{Noise2Music Waveform} & \textbf{MusicLM} & \textbf{Mubert} & \textbf{Riffusion}\\
    \midrule
     959 & 718 & 692 & 254 & 308\\
    \bottomrule
  \end{tabular}}
\vskip-0.15in
\end{table}

\subsection{Inference parameter ablations}

We vary inference parameters of the models and observe its effects. We note that we have conducted ablations with model checkpoints that are slightly less-trained compared to the checkpoints used to produce the evaluation numbers in the previous subsection. The ablations are conducted with respect to the base parameters as listed in \cref{t:model-inference-params}.

\begin{figure}[ht]
\vskip 0.05in
\begin{center}
\centerline{\includegraphics[width=0.7\columnwidth]{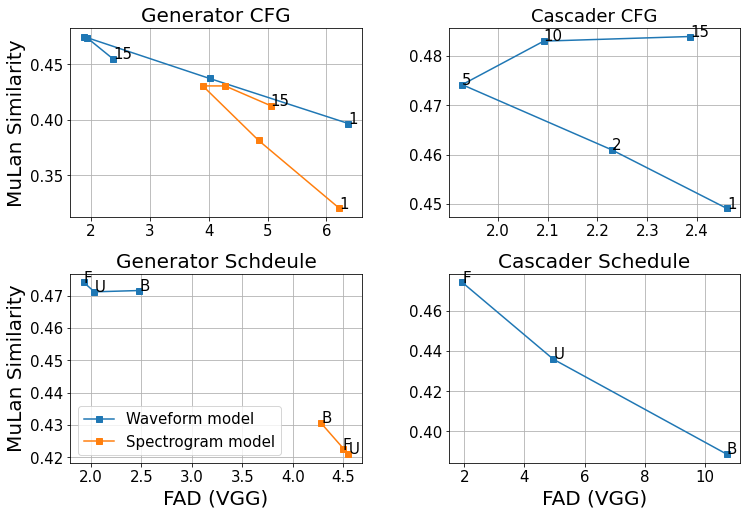}}
\vskip -0.05in
\caption{We plot how $\text{FAD}_\text{VGG}$ and the MuLan similarity score vary as inference parameters are adjusted. The CFG parameters take values from [1, 2, 5, 10, 15], while ``B"ack-heavy, ``U"niform and ``F"ront-heavy denoising step schedules have been applied.}
\label{f:ablations}
\end{center}
\vskip -0.25in
\end{figure}

In \cref{f:ablations}, we depict how FAD measured with VGG and the MuLan similarity score change as the denoising step schedule and the CFG scale are varied during inference. Only one parameter is varied at a time, and all other parameters stay fixed at the baseline values.

We find an overall correlation between the FAD metric and the similarity score, except in the case of the cascader, where FAD can get worse while the similarity score improves. We also find that there is an optimal CFG scale, and too big of a CFG scale hurts the generation quality. It can also be seen that the generator CFG scale is a bigger factor than the denoising schedule of the generator, while the impact of cascader denoising schedule is extremely large.

\begin{figure}[ht]
\begin{center}
\centerline{\includegraphics[width=0.7\columnwidth]{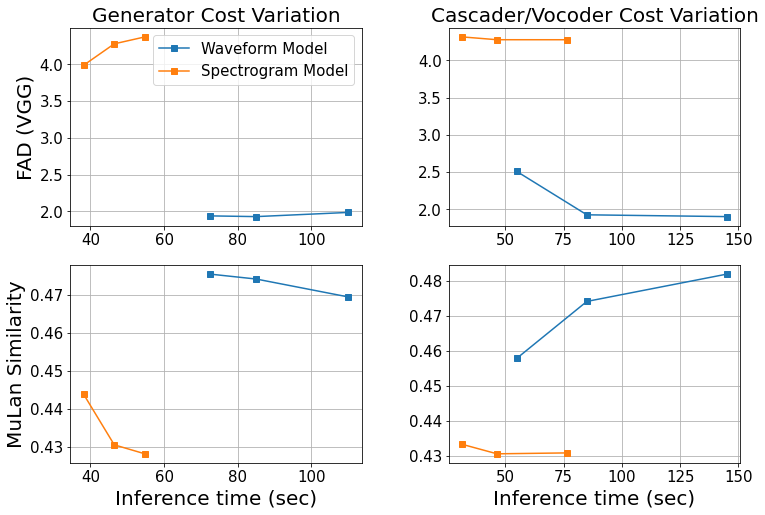}}
\caption{Quality metrics of the generated examples plotted against computational cost parameterized by inference time.}
\label{f:cost}
\end{center}
\vskip -0.25in
\end{figure}

\subsection{Inference cost and performance}

In \cref{f:cost}, we plot the quality metrics against the inference cost measured by the inference time. We reduce/increase the number of inference steps of the generator or the cascader/vocoder and inverse-proportionally scale the step sizes in the inference schedule. We find that the effect of increasing the inference cost of the generator is mixed while the generative quality generally improves with more cascader/vocoder inference steps.

\section{Qualitative analysis}

\textbf{Content representation:}  We present generation examples at \href{\website#table-2}{\websiteDisplay\#table-2}, to illustrate that the model is able to ground the music aspects represented in the text prompt. In particular, we find that the genre, instrument, mood, vocal traits, and era of music implied in the text is manifested in the generated music.

\textbf{Creative prompts:} While our models often struggle to produce high quality audio from out-of-distribution prompts, they are nevertheless able to generate some interesting examples. In \href{\website#table-3}{\websiteDisplay\#table-3}, we have collected examples of creative prompts for which the model was able to generate quality music.

\section{Discussion}

\textbf{Spectrogram vs. waveform approach:} The spectrogram and waveform approaches have their comparative advantages. The spectrogram models employed in this work are much cheaper to train and serve compared to the waveform models, and are more scalable in time length. This is because the sequence length of the spectrogram is much shorter than that of a low-fi waveform. In addition, the spectrogram contains high-frequency information which is missing in the low-fidelity audio. Meanwhile, the waveform model produces interpretable representations at every step of the generation process, making the model easy to debug and tune. This is partially responsible for our ability to train the waveform models with more ease.

\textbf{Future directions:} While we have demonstrated the potential of text prompt based music generation, there is much room for improvement beyond our work. Increasing model interpretability, further improving text-audio alignment, reducing training and inference cost, and scaling up the generation length of the audio are just a few directions in which our work needs to be improved. Another interesting direction is to fine-tune the models trained in this work for diverse audio tasks including music completion and modification, as was done for images by \citet{saharia2022palette}.

\section{Broader Impact}

We believe our work has the potential to grow into a useful tool for artists and content creators that can further enrich their creative pursuits. To live up to this promise, more work is needed with musicians and other stakeholders to develop models into a meaningful co-creation tool. 

We acknowledge the limitations of the proposed model. In particular, large generative models learn to imitate patterns and biases inherent in the training sets, and in our case, the model can propagate the potential biases built in the text and music corpora used to train our models. Such biases can be hard to detect as they manifest in often subtle, unpredictable ways, which are not fully captured by our current evaluation benchmarks. Demeaning or other harmful language may be generated in model outputs, due to learned associations or by chance.

Beyond this, we recognize that musical genres are complex and key musical attributes are contextual and change over time. Training data reflect a limited corpus of musical samples and genres, given uneven recording and digitization of samples from global musical cultures. How music is categorized and labeled can essentialize genres; and these labels may be constructed and applied without the participation of communities. 
When readers examine the released generation examples in the accompanied website, we caution readers not to presume each sample can generalize to an entire musical genre or one label can capture the diversity of musical genres produced within a region (i.e. ``Latin music" contains a broad range of cultures and styles). Moreover, musical samples may sound ``authentic" to those outside these communities, as nuances in musical traditions need trained ears/cultural knowledge to recognize. In generating vocals, there may be possible caricatures, 11mock accents," parodies, or other demeaning linguistic harms (e.g., ``mock Black singing" in a request for ``soulful vocals" or ``mock Spanish" in a Latin music request) that arise in text prompts requesting cultural or religious musical genres, or genres that emerged as part of the political struggles of certain communities (e.g., Black American music, Nueva canci\'on, Chicano folk, Brazilian Tropicalismo, Sufi Qaw).

As is with any other technology, the result of our research can be misused or abused. We acknowledge the risk of potential misappropriation when the created content exactly matches examples in training data. In accordance with responsible model development practices, duplication checks are a built-in part of our current pipeline of producing and releasing examples, and will continue to be for any future work.

Efforts for identifying potential safety issues and addressing them are important components for improving these generative models. Until there is a more clear understanding of the limitations and risks, we do not intend to release the model.

\section*{Acknowledgements}
We are grateful to Aren Jansen for building MuLan, which is an indispensable component of this project. We give thanks to Austin Tarango, Fernando Diaz, Kathy Meier-Hellstern, Molly FitzMorris, and Renee Shelby for helping us incorporate important responsible AI practices around this project. We acknowledge support from Blake Cunningham, Cara Adams, for giving us advice along the project and assisting us with the publication process. We appreciate valuable feedback and support from Alex Ku, Andrea Agostinelli, Ankur Bapna, Chen Liang, Ed Chi, Ekin Dogus Cubuk, Erica Moreira, Esteban Real, Heiga Zen, Jaehoon Lee, James Qin, Nathan Park, Stephen Kelly, Thang Luoung, Weizhe Hua, Ye Jia, Yifeng Lu, Yonghui Wu, Yu Zhang, Yuma Koizumi. Special thanks to authors of MusicLM for helpful discussions and cooperation, and especially for sharing their evaluation set and manuscript before publication.

\bibliography{main}
\bibliographystyle{icml2023}

\newpage
\appendix
\onecolumn

\section{Diffusion models}
\label{app:diffusion}

In this section, we review some relevant information for diffusion models and set up the notation used in the main text of the paper. We follow \cite{saharia2022photorealistic} in our presentation.

A diffusion model assumes a set-up where a sample $\mathbf{x}$ from a distribution corrupted by a Gaussian diffusion process with a noise schedule, represented by a monotonically increasing standard deviation $\sigma_t$ at time $t$. More precisely, the distribution for the corrupted sample $\mathbf{x}_t$ at time $t$ conditioned on $\mathbf{x}$ or $\mathbf{x}_s$ for $s < t$ is given by:

\begin{equation}
q(\mathbf{x}_t |\mathbf{x}) = \mathcal{N}(\alpha_t \mathbf{x}, \sigma_t^2 \mathbf{I}) \,, \qquad
q(\mathbf{x}_t |\mathbf{x}_s) = \mathcal{N}((\alpha_t/\alpha_s) \mathbf{x}_s, \sigma_{t|s}^2 \mathbf{I}) \,.
\end{equation}
We can define a set of variables that will make the equations cleaner:
\begin{equation}
\alpha_t = \sqrt{1 - \sigma_t^2} \,, \quad
\lambda_t = \ln(\alpha_t^2/\sigma_t^2) \,, \quad
\sigma_{t|s}^2 = (1 - e^{\lambda_t - \lambda_s}) \sigma_t^2 \,, \quad
\tilde{\sigma}_{s|t}^2 = (1 - e^{\lambda_t - \lambda_s}) \sigma_s^2 \,.
\end{equation}
The time variables $s$ and $t$ are assumed to be in the range $[0, 1]$.

As seen in the first equation, the randomness of the corrupted sample $\mathbf{x}_t$ is encoded in a single noise vector $\epsilon \sim \mathcal{N}(0, \mathbf{I})$. The aim of the diffusion model is to model this noise vector, given the corrupted sample, the time $t$ and the context $\mathbf{c}$: $\bm{\epsilon}_\theta(\mathbf{x}_t, \mathbf{c}, t)$.

Once the model is trained, we sample noise at time $t=1$, i.e., $x_1 \sim \mathcal{N}(0, \mathbf{I})$, and reverse the diffusion process to produce a ``clean" sample $\mathbf{x}_0$ from the original distribution. We employ ancestral (or DDPM) sampling \cite{ho2020denoising} to do so. In this sampling method, we select some time steps $0 = t_0 < \cdots < t_N = 1$ and reverse the diffusion process by applying the update rule to obtain $\mathbf{x}_s$ from $\mathbf{x}_t$:
\begin{equation}
\mathbf{x}_s = {\alpha_s \over \alpha_t} \mathbf{x}_t -
(1 - e^{\lambda_t - \lambda_s})\cdot {\alpha_s \over \alpha_t} \cdot \sigma_t \cdot \bm\epsilon_\theta(\mathbf{x}_t, \mathbf{c}, t) +
\tilde{\sigma}_{s|t}^{1 - \gamma} \cdot {\sigma}_{t|s}^{\gamma} \cdot \bm{\tilde{\epsilon}} \,,
\end{equation}
where $s = t_{N-n}$ and $t = t_{N-n+1}$ at the $n$-th update step. Here, $\bm{\tilde{\epsilon}}$ is a random standard normal vector sampled at each inference step. $\gamma$ is a hyperparameter that controls the stochasticity of the diffusion process. As $\gamma$ increases, a larger variance is introduced at a given inference step since $\sigma_{t|s} > \tilde{\sigma}_{s|t}$ for $t > s$.

\subsection{Noise schedules}

We utilize two noise schedules---the linear \cite{ho2020denoising} and cosine \cite{nichol2021improved} schedules.

The linear schedule \cite{ho2020denoising} is defined to be such that $\sigma_t^2$ increases in a linear fashion with respect to time:
\begin{equation}
\sigma_t^2 = (\sigma_1^2 - \sigma_0^2) t + \sigma_0^2 \,.
\end{equation}
We use the initial and final values $\sigma_0^2 = 0.0001$ and $\sigma_1^2 = 0.02$ in this work.

We use a slight variant of the cosine schedule \cite{nichol2021improved}:
\begin{equation}
\alpha_t = \cos (at + b)
\end{equation}
where $a=\arctan e^{10} - \arctan e^{-10}$ and $b=\arctan e^{-10}$ are taken so that $\lambda_{t=0} = 20$ and $\lambda_{t=1} = -20$.

\subsection{Denoising step schedules}

\begin{figure}[h!]
\begin{center}
\centerline{
\includegraphics[width=0.7\columnwidth]{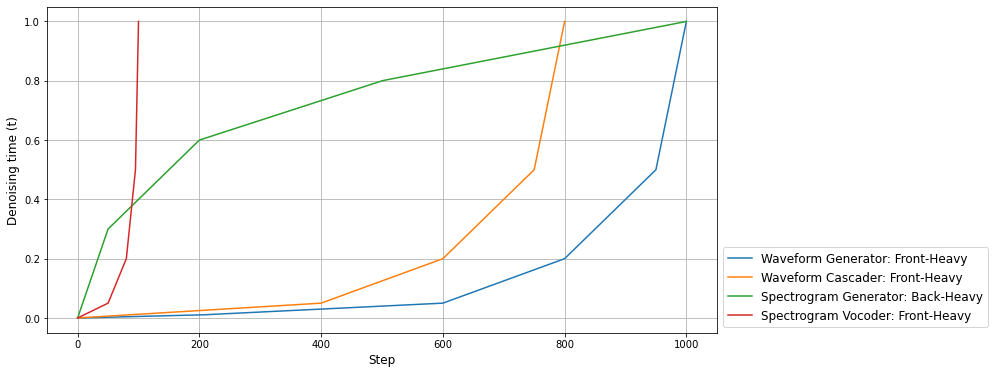}}
\caption{The denoising time reached at a given inference step for the four denoising time step schedules used in the paper. A front-heavy schedule expends most of its steps near $t=0$ while a back-heavy schedule expends most of its steps near $t=1$.}
\label{figure:dn}
\end{center}
\vskip -0.2in
\end{figure}

Here we list the actual denoising step schedules used at inference time for the generator and cascader/vocoder models. These schedules have been hand-tuned by trial and error as explained in section \ref{ss:model-parameter-selection}. The numerical values of the time step sizes for each model and schedule used for ablation is given in python notation in \cref{table:denoising-schedules}. The schedules used for final evaluation are italicized. Notice for the spectrogram generator, the back-heavy schedule has been modified further for evaluation. Meanwhile, a plot of the denoising time reached at a given inference step for each model schedule is plotted in \cref{figure:dn}.

\begin{table}[h!]
\caption{The denoising time steps values for denoising schedules. The schedules selected for use for each model are italicized.}
  \label{table:denoising-schedules}
  \vskip 0.05in
  \centering
  \resizebox{0.95\columnwidth}{!}{\begin{tabular}{lcl}
    \toprule
    \bf Model & \bf Schedule & \bf Time steps\\
    \midrule
    Waveform generator
    & \it Front-heavy & \texttt{front\_heavy = [0.01 / 200] * 200 + [0.04 / 400] * 400 + [0.15 / 200] * 200 + [0.3 / 150] * 150 + [0.5 / 50] * 50} \\
    & Back-heavy & \texttt{back\_heavy = front\_heavy[::-1]} \\
    & Uniform & \texttt{uniform = [1.0 / 1000] * 1000} \\
    \midrule
    Waveform cascader
    & \it Front-heavy & \texttt{front\_heavy = [0.05 / 400] * 400 + [0.15 / 200] * 200 + [0.3 / 150] * 150 + [0.5 / 50] * 50} \\
    & Back-heavy & \texttt{back\_heavy = front\_heavy[::-1]} \\
    & Uniform & \texttt{uniform = [1.0 / 800] * 800} \\
    \midrule
    Spectrogram generator
    & Front-heavy & \texttt{front\_heavy = [0.01 / 400] * 400 + [0.04 / 800] * 800 + [0.15 / 400] * 400 + [0.3 / 300] * 300 + [0.5 / 100] * 100} \\
    & Back-heavy & \texttt{back\_heavy = front\_heavy[::-1]} \\
    & Uniform & \texttt{uniform = [1.0 / 1000] * 1000} \\
    \cmidrule{2-3}
    & \it Back-heavy & \texttt{back\_heavy\_eval = [0.3 / 50] * 50 + [0.3 / 150] * 150 + [0.2 / 300] * 300 + [0.2 / 500] * 500}\\
    \midrule
    Spectrogram vocoder
    & \it Front-heavy & \texttt{front\_heavy = [0.05 / 50] * 50 + [0.15 / 30] * 30 + [0.3 / 15] * 15 + [0.5 / 5] * 5} \\
    & Back-heavy & \texttt{back\_heavy = front\_heavy[::-1]} \\
    & Uniform & \texttt{uniform = [1.0 / 100] * 100} \\
    \midrule
    Super-resolution cascader
    & \it Front-heavy & \texttt{front\_heavy = [0.05 / 400] * 400 + [0.15 / 200] * 200 + [0.3 / 150] * 150 + [0.5 / 50] * 50} \\
    \bottomrule
  \end{tabular}}
\end{table}

\section{Model architecture}
\label{app:architecture}

We present some details on the architecture of the models used in the paper. The overall structure of the U-Net has been depicted in \cref{U-Net}. After an entry convolutional layer is applied to the input, the input is passed through a series of down-sampling and up-sampling layers with convolutional blocks in between. Here we expand upon how the convolutional blocks in the down-sampling/up-sampling portion of the U-Nets are structured, and explain how each network utilized in the generation pipeline is configured.

\subsection{Down-sampling and up-sampling blocks}

The model architecture closely follows that of the efficient U-Net \cite{saharia2022photorealistic}, with two-dimensional convolutional layers replaced by one-dimensional convolutional layers. There are small differences, which we review here.

As seen in \cref{U-Net}, the U-Net model is a mirror image of itself. Given the model depth $D$, the model employs $D$ down-sampling and $D$ up-sampling layers, which we may label by $I = 1 , \cdots, D$. Each down-sampling layer is a one-dimensional convolutional layer with stride $S_I$. Thus, denoting the feature length and feature dimension of the forward-propagated input at depth-$I$ to be $T_I$ and $C_I$, the down-sampling layer takes an input with (length, channel) dimensions $(T_{I-1}, C_{I-1})$ and maps it to an output with dimensions $(T_I, C_I)$ with $T_I = T_{I-1}/S_I$. The up-sampling layer does the mirror operation, where an input with $(T_I, C_I)$ dimensions is mapped to an output with $(T_{I-1}, C_{I-1})$ dimensions, with up-sampling stride $S_I$.

\begin{figure}[h!]
\begin{center}
\centerline{
\includegraphics[width=0.5\columnwidth]{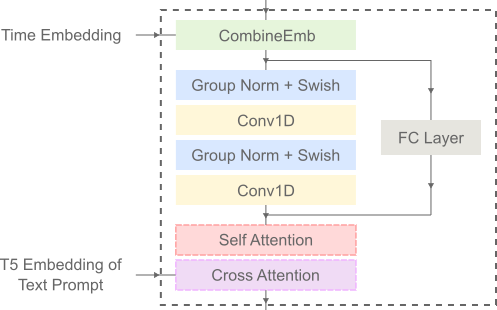}}
\caption{The structure of the convolutional blocks that form the base unit of operation in the 1D U-Nets. The self and cross attention layers are optional.}
\label{figure:cblock}
\end{center}
\vskip -0.2in
\end{figure}

Between the down-sampling and up-sampling layers, ``convolutional blocks" of uniform dimension are used. The exact same block is used for both the down-sampling and the up-sampling portion of the U-Net, and the same number of blocks are employed at the same depth. The structure of a block is depicted in \cref{figure:cblock}. While all blocks interact with the time embedding vector---obtained by converting the float into a vector via positional embedding and applying a linear layer---through the ``combine embedding" layer, the self attention and the cross attention layers are only turned on for selected depths, or not even used at all in some cases. The ``combine embedding" layer applies a fully connected layer to the time embedding to compute a channel-wise scaling and bias vector, which is applied to the input sequence. The self and cross attention layers include the standard post-attention residual layer with a hidden layer of twice the dimension of $C_I$.

Meanwhile, the entry convolutional layer and the exit convolutional layer both have kernel size 3. The ``zeroth" channel width of the input to the first down-sampling layer and the output of the last up-sampling layer are both set to a ``base model dimension". The time embedding dimension is also set to this value.

Let us now summarize the key hyperparameters that determine the architecture of the network.
\begin{itemize}
\item Base model dimension
\item Convolutional kernel size
\item Depth of the network $D$
\item Length-$D$ list of down-sampling factors
\item Length-$D$ list of number of blocks used at each depth
\item Length-$D$ list of bools indicating whether self/cross-attention is used at a given depth
\item Number of heads used for attention
\end{itemize}

The rest of the architecture is fixed. 

\subsection{Model specifications}

The architectural parameters of the models used in this paper can be summarized by \cref{table:architecture-parameters}.

\begin{table}[h!]
\caption{Architectural parameters for the models used in this work.}
  \label{table:architecture-parameters}
  \vskip 0.05in
  \centering
  \resizebox{0.95\columnwidth}{!}{\begin{tabular}{lrrrrrr}
    \toprule
    \bf Parameter &
    & \bf Waveform generator 
    & \bf Waveform cascader
    & \bf Spectrogram generator 
    & \bf Spectrogram vocoder
    & \bf Super-resolution cascader\\
    \midrule
    Base dimension &
    & 256
    & 256 
    & 256
    & 128
    & 256 \\
    Kernel Size &
    & 7
    & 7 
    & 9
    & 5
    & 7 \\
    Depth &
    & 6
    & 5
    & 5
    & 4 
    & 4\\
    Strides &
    & [4, 4, 4, 4, 4, 4]
    & [4, 4, 4, 4, 4]
    & [4, 3, 2, 2, 2]
    & [2, 4, 4, 5] 
    & [4, 4, 4, 4] \\
    \# Channels &
    & 256 * [1, 1, 2, 3, 4, 4] 
    & 256 * [1, 1, 2, 3, 4] 
    & 256 * [1, 2, 2, 3, 4]
    & 128 * [1, 1, 2, 4] 
    & 256 * [1, 1, 2, 2] \\
    \# Blocks &
    & [3, 3, 3, 4, 4, 4]
    & [3, 3, 4, 6, 6] 
    & [6, 6, 6, 6, 6]
    & [4, 4, 4, 4]
    & [4, 4, 4, 4] \\
    Self Attention &
    & [F, F, F, T, T, T]
    & [F, F, F, F, F]
    & [F, F, F, T, T]
    & [F, F, F, F]
    & [F, F, F, F] \\
    Cross Attention &
    & [F, F, F, T, T, T]
    & [F, F, F, T, T]
    & [F, F, F, T, T]
    & [F, F, F, F]
    & [F, F, F, F] \\
    Attention Heads &
    & 8
    & 8
    & 8
    & - 
    & - \\
    \bottomrule
  \end{tabular}}
\end{table}

\section{Prompt template to prime LaMDA model to generate music descriptive text}
\label{app:ladma-prompt-template}

``Walking on Sunshine" by Katrina \& The Waves :
The song is a pop / rock song. It has a happy, upbeat mood, with a driving bassline and a simple, repetitive drumbeat. The song is backed by a synthesizer and a guitar. 

``Born This Way" by Lady Gaga :
The electropop song is backed by rumbling synth sounds, a humming bass and additional chorus percussion, with sole organ toward the end. The female singer has a confident and strong voice. 

``Till I Collapse" by Eminem :
The song is a hip hop song. It has a strong, aggressive mood, with a powerful, energetic beat. 

``My Heart Will Go On" by Celine Dion :
The pop song has a sad and heartbroken mood. It contains heavy emphasis on the instrumental arranging. Usage of Tin Whistle is prominent, backed by melodic use of strings and rhythm guitars. The song features both acoustic and electronic instrumentation. The female singer's vocal performance is emotional demanding. 

``Me Too" by Meghan Trainor :
The pop song is a mid-tempo pop song. It has a happy, upbeat mood, with a driving bassline and a simple, repetitive drumbeat. The song is backed by a synthesizer and a guitar. 

``Nuvole Bianche" by Einaudi :
The song is a piano-based classical piece. It has a gentle, melancholy mood, with a soothing, slow pace. 

``Don't Stop Me Now" by Queen :
The energetic rock song builds on a piano, bass guitar, and drums. The singers are excited, ready to go, and uplifting. 

``Strawberry Swing" by Coldplay :
The alternative/indie song contains influences from afro-pop and highlife music, and is built around finger-picked, distortion-free guitars with a heavy bassline and psychedelic synths. It's a mid-tempo track, featuring echoing guitars, piano ballad-inspired melodies and bittersweet, anthemic falsetto vocals. 

``Mad World" by Gary Jules :
The new wave / synth-pop song is backing its male singer with only a set of piano chords, a mellotron imitating a cello, very light touches of electric piano, and modest use of a vocoder on the chorus. 

``A Change is Gonna Come" by Sam Cooke :
The soul / R\&B song has a reflective, nostalgic mood. The male singer's voice is clearly in the foreground, backed by horns, strings, and the timpani carrying the bridge. The French horn conveys a sense of melancholy. 

``Smells Like Teen Spirit" by Nirvana :
The alternative rock/hard rock song has quiet verses with wobbly, chorused guitar, followed by big, loud hardcore-inspired choruses. The overall mood is rebellious and pumped up. The guitar chords are double tracked to create a more powerful sound. 

``Strawberry Fields Forever" by The Beatles :
The psychadelic rock song features a reverse-recorded instrumentation, Mellotron flute sounds, an Indian swarmandal, tape loops and a fade-out/fade-in coda, as well as a cello and brass arrangement. The vocals are slightly dissonant adding a bittersweet and ominous quality. 

\{title\} by \{artist\} :

\section{AudioSet music labels}
\label{app:audioset-music-tags}

AudioSet labels are licensed under the Creative Commons Attribution-ShareAlike 4.0 International (CC BY-SA 4.0) license.

/m/0z9c,/m/0mkg,/m/042v\_gx,/m/0fd3y,/t/dd00036,/m/025td0t,/m/0192l,/m/018j2,/m/0bm02,/m/018vs,/m/02cz\_7,/m/0395lw,

/m/0gg8l,/m/0155w,/m/0l14\_3,/m/01kcd,/m/015vgc,/m/01xqw,/m/02bk07,/m/0l14jd,/m/02mscn,/m/0140xf,/m/01wy6,/m/0ggq0m,

/m/01lyv,/m/0239kh,/m/01qbl,/m/0ggx5q,/m/02bxd,/m/026z9,/m/02fsn,/m/0283d,/m/02hnl,/m/02k\_mr,/m/026t6,/m/07s72n,

/m/02sgy,/m/08cyft,/m/02lkt,/m/03xq\_f,/m/0m0jc,/t/dd00035,/m/0326g,/m/0l14j\_,/m/02w4v,/m/0319l,/m/02x8m,/t/dd00032,

/m/0dwtp,/m/0mbct,/m/0dls3,/m/0342h,/m/03gvt,/t/dd00031,/m/03qjg,/m/03m5k,/m/03q5t,/m/03lty,/m/0glt670,/m/03mb9,

/m/05rwpb,/m/03\_d0,/m/03r5q\_,/m/05148p4,/m/07pkxdp,/m/0j45pbj,/m/04rzd,/m/0dwsp,/m/06j64v,/m/05fw6t,/m/0164x2,

/m/028sqc,/m/0dq0md,/m/0g293,/m/02v2lh,/m/05pd6,/m/013y1f,/m/0l14md,/m/05r5c,/m/0fx80y,/m/064t9,/m/0dl5d,/m/05w3f,

/m/05r6t,/m/05r5wn,/m/06cqb,/m/06j6l,/m/03t3fj,/m/07sbbz2,/m/06by7,/t/dd00033,/m/0ln16,/m/06ncr,/t/dd00037,

/m/01hgjl,/m/0l14l2,/m/0l14t7,/m/0jtg0,/m/06rqw,/m/06rvn,/m/0gywn,/m/0l14gg,/m/06w87,/m/0l156b,/m/02qmj0d,

/m/07s0s5r,/m/015y\_n,/m/0l14qv,/m/01p970,/m/07brj,/m/01glhc,/m/07gxw,/t/dd00034,/m/02cjck,/m/07kc\_,/m/011k\_j,

/m/02p0sh1,/m/07lnk,/m/07c6l,/m/07gql,/m/016622,/m/07xzm,/m/0dwt5,/m/01z7dr,/m/07y\_7,/m/0y4f8,/m/04wptg,/m/085jw,

/m/01sm1g,/m/01bns\_

\end{document}